Opinion paper

Critical rationalism and the search for standard

(field-normalized) indicators in bibliometrics

Lutz Bornmann* & Werner Marx**

*Division for Science and Innovation Studies

Administrative Headquarters of the Max Planck Society

Hofgartenstr. 8,

80539 Munich, Germany.

E-mail: bornmann@gv.mpg.de

**Max Planck Institute for Solid State Research

Information Service

Heisenbergstrasse 1,

70506 Stuttgart, Germany.

Email: w.marx@fkf.mpg.de

**Abstract**

Bibliometrics plays an increasingly important role in research evaluation. However, no gold standard exists for a set of reliable and valid (field-normalized) impact indicators in research evaluation. This opinion paper recommends that bibliometricians develop and analyze these impact indicators against the backdrop of Popper's critical rationalism. The studies critically investigating the indicators should publish the results in such a way that they can be included in meta-analyses. The results of meta-analyses give guidance on which indicators can then be part of a set of indicators used as standard in bibliometrics. The generation and continuous revision of the standard set could be handled by the International Society for Informetrics and Scientometrics (ISSI).

# 1      Introduction

Bibliometrics plays an increasingly important role in research evaluation (Bornmann, 2017). An evaluation report without bibliometrics is currently difficult to imagine. In the application of bibliometrics in evaluative practice, one can differentiate between professional and citizen bibliometrics (Leydesdorff, Wouters, & Bornmann, 2016). Citizen bibliometrics is based on simple indicators, such as the median and total number of citations. These indicators can possibly be used to compare similar units (concerning size and age) in the same or very similar fields. An important disadvantage of these indicators is, however, that they cannot be used in cross-field comparisons of units (e.g. research groups or institutions). The publication and citation cultures are so different in the fields that the median and total number of citations from different fields are not comparable.

In professional bibliometrics, established field-normalized indicators are applied in cross-field comparisons, which take into account that citations occur in different publication and citation cultures (Waltman, 2016). Normalized indicators can be used to assess whether a unit (e.g. an institute) shows above or below average performance in the corresponding fields. Most of the studies in bibliometrics – scientific studies and applied studies used to support research evaluations – are actually cross-field comparisons, because bibliometrics is especially suitable for greater units (e.g. institutes or countries) and these units publish as a rule in different fields. We argue in this opinion paper that bibliometrics should aim to develop standard approaches especially to field-normalization. Only with standard approaches, the results of different studies are comparable, reliable, and interpretable.

In bibliometrics, however, several variants of normalized indicators exist. For example, mean-based indicators compare the mean impact of the focal papers with the mean impact of papers published in the same field and publication year. Percentile-based indicators assign to every paper in a certain field and publication year a citation percentile, which can be used for cross-field comparisons. Some bibliometricians recommend mean-based citation indicators (e.g., the characteristic scores and scales method; see Glänzel & Schubert, 1988) and others deem percentiles the best alternative among the indicators (Bornmann, Leydesdorff, & Mutz, 2013). Further, different variants of mean-based and percentile-based indicators are currently in use (Todeschini & Baccini, 2016). No gold standards exist which can be recommended for the general application of impact indicators in research evaluation.

If it should be the aim of bibliometrics to develop indicators which gain general acceptance (among bibliometricians and beyond), it is essential to criticize the available indicators sharply. This is especially necessary for field-normalized indicators, because these indicators have – as a rule – a more complicated design than simple indicators such as mean citation rates. We propose that indicators should be continuously investigated as to their reliability and validity against the backdrop of Popper's critical rationalism (Gingras, 2016; Lindner, Torralba, & Khan, 2018; Ulrich, 2006): bibliometricians should try to create or find situations in which the indicators possibly do not show the desired behavior. For example, an indicator which is intended to reflect the quality of publications should be correlated with the assessments of peers (Bornmann & Marx, 2015). If the correlation turns out to be low, the validity of the indicator can be questioned. Situations should be created or found in empirical research in which the measurement with the new indicator can fail. An indicator is regarded as preliminarily valid if these situations could not be found or realized. Such an indicator should gain general acceptance in citizen and professional bibliometrics.

In bibliometrics, we frequently find inadequate investigations of the validity of indicators in the literature. Good examples are the developments of h index variants since the introduction of the h index in 2005 (Hirsch, 2005). Since then, about 50 variants have been introduced (Bornmann, Mutz, Hug, & Daniel, 2011). In many of the corresponding papers, one or several weaknesses of the h index are unrevealed, before the new variant is introduced (see, e.g., Tol, 2009). Then, it is demonstrated on the base of one or several examples that the new variant performs (significantly) better than the original index. However, these tests do not guarantee that the indicators have been proved successfully. The authors seem to be more interested in producing favorable results for the new indicators than their empirical validation. For critical validation, situations should be sought in which the indicators probably fail and these situations should be specifically investigated. With the elimination of failed indicators, the set of indicators is reduced and standard indicators emerge.

For the critical examination of (field-normalized) indicators, some empirical and theoretical tests have been proposed which are explained in the following two sections. Since the success of field-normalization depends on the used field categorization scheme, we discuss the current (most frequently) used schemes in section 4. Section 5 introduces the Werther effect describing the situation in which indicators become popular although theoretical and empirical results question their use.

## 2    Empirical tests of field-normalized indicators

For the examination of the convergent validity, the correlation between the field-normalized indicator and the assessment by peers is calculated (Bornmann & Marx, 2015). The backdrop of the test is that both methods of research evaluation measure the same construct, namely the quality of research (Thelwall, 2017). The peer review process is the oldest and most accepted method for research evaluation (Bornmann, 2011). If the correlation coefficient is substantial, the field-normalized indicator seems to measure research quality validly. The substantial coefficient should be at least at the medium level (about $r=.5$). We cannot expect a high correlation (about $r=.7$), since citations only measure one aspect of quality, namely impact. Importance, for instance, can scarcely be measured by citations (Martin & Irvine, 1983). The coefficient ($r=.5$) which should be reached by the correlation between field-normalized indicators and assessments by peers is a rough estimation, since the correlation can be performed on different levels of aggregation: correlations will usually be lower at the level of individual publications than at the level of research groups, departments, or research institutes.

At least three further tests have been proposed which investigate the ability of indicators to minimize the effect of field-differences in comparison (e.g. mean-based citation indicators compared to percentile indicators). A valid field-normalized indicator adjusts the citation impact of a paper to the average field level and allows meaningful cross-field comparisons. The three tests require a field categorization scheme (see section 4) which is independent of the scheme used for field normalization (Sirtes, 2012; Waltman & van Eck, 2013). This is also the disadvantage of these tests: one needs a second independent subject category scheme, which is able to reflect scientific fields properly. If such a scheme is not available, the tests cannot be applied.

The first test can be called "fairness test" (Bornmann, de Moya Anegón, & Mutz, 2013; Kaur, Radicchi, & Menczer, 2013; Radicchi, Fortunato, & Castellano, 2008). According to Radicchi and Castellano (2012) "the 'fairness' of a citation indicator is … directly quantifiable by looking at the ability of the indicator to suppress any potential citation bias related to the classification of papers in disciplines or topics of research" (p. 125). For example, Bornmann and Haunschild (2016a, 2016b) used the fairness test to study field-normalized indicators in the area of bibliometrics (for the testing of the indicator: citation score normalized by cited references, CSNCR) and altmetrics (Mendeley reader indicators). Other applications have been reported in the bibliometric literature.

In the first step of the fairness test, all papers from one publication year are sorted in descending order by the normalized indicator. Then, the 10% of papers with the highest indicator values are identified and a new binary variable is generated, where 1 marks highly cited papers and 0 the rest (90% of the papers). In the second step of the test, all papers are grouped by the subject categories of the independent field-categorization scheme. In the third step of the fairness test, the proportion of papers belonging to the 10% of papers with the highest indicator values is calculated for each subject category – using the binary variable from the first step. An indicator can be called fair if the proportion of highly cited papers within the subject categories equals the expected value of 10%. In other words, the smaller the deviations from 10% are, the less the indicator is dependent on specific fields. The fairness test should also consider bare citation counts. The field-normalized indicator should show smaller deviations from 10% than bare citation counts.

In the second test, the field-normalized values are grouped by the subject categories of the independent field-categorization scheme. Then, the intra-class correlation coefficient (ICC) – an inferential statistic – can be used to investigate how strongly the values in the same subject category resemble each other (Leydesdorff & Bornmann, 2011). Within the framework of random effects models, the ICC can be interpreted as the proportion of total variance between the subject categories. This variance proportion should be as small as possible for a valid field-normalizing procedure. At least, the variance proportion should be smaller than that for bare citation counts. A proposed new field-normalizing procedure should perform at least on the same level as already established indicators, such as the mean-based citation indicators or percentiles. Meaningful differences between ICC values for field-normalized indicators can be obtained by calculating confidence intervals.

Li and Ruiz-Castillo (2013) introduced graphical and numerical evaluation approaches which can be used to study field-normalized indicators – the third test. The approaches are based on a similar solution as the second test (by comparing within- and between group differences). The measurement framework of the approaches is constituted by Crespo, Li, and Ruiz-Castillo (2013). In this framework, differences in citation cultures across fields are captured by a between-group term: "by using an additively decomposable inequality index, in which case the citation inequality attributed to differences in citation practices is captured by a between-group inequality term in the double partition by field and citation quantile" (Crespo et al., 2013). Perianes-Rodriguez and Ruiz-Castillo (2017) used the framework to study different field-categorization systems on different granularity levels. Based on the empirical results, they recommend, for instance, to use "the system at the higher level because it typically exhibits a better standard normalization performance" (p. 43).

Individual studies investigating field-normalized indicators should be designed and presented in such a way that the results can be included in meta-analyses. Meta-analyses summarize quantitatively the results of single studies in order to arrive at a generalized statement across all studies. For example, Bornmann (2015) has published a meta-analysis of studies which have investigated the correlation between citations and altmetrics. He was interested in the relationship between the traditional method of measuring impact (by using citations) and the new ways (by using altmetrics). The results show, e.g., a very low correlation in general for Twitter counts. In another meta-analysis, Bornmann et al. (2011) undertook an analysis of studies investigating the correlation between the h index and h index variants. They wanted to know whether the variants measure similar or different things compared with the original h index. The generally high correlation between the h index and its variants demonstrates the redundancy of these variants. In a similar way as in these published meta-analyses, meta-analyses could be used in the assessment of field-normalized indicators to arrive at conclusions on the basis of the results of individual empirical studies. Field-normalized indicators which show favorable results in meta-analyses should be preferred in research evaluation. For example, indicators showing the highest pooled correlation with the assessments by peers should be preferred.

Since it is relatively easy to carry out large-scale empirical analyses in bibliometrics, an alternative to meta-analysis could be to have one or a few large-scale analyses including the data from different databases (e.g. Web of Science, Scopus, Google Scholar, Microsoft Academic) instead of a meta-analysis including a large number of small-scale analyses. The large-scale studies should be ideally undertaken in collaboration by a group of bibliometricians working on the same set of indicators to promote general acceptance of the study results.

## 3      Mathematical and statistical tests of field-normalized indicators

Empirical studies investigating field-normalized indicators are as a rule sophisticated and need access to citation indexes and other databases. As an alternative (or additional) method of investigation, the statistical and mathematical properties of the indicators can be tested. Some of these properties are so basic and natural that one can expect them from indicators. A good example is the use of measures of central tendency to identify the central position within citation data. Since citation data are as a rule skewed data, other measures of central tendency than the arithmetic average should be used (especially the median). The arithmetic average is inappropriate if values are unusual – being especially small or large – compared to the rest of the data set. Since mean-based citation indicators use the arithmetic average in their definition, they are actually scarcely appropriate for field-normalizing citations, and alternative indicators should be used. Although many bibliometricians (including the authors of this opinion paper) are aware of this problem, they use mean-based citation indicators, because they are (still) standard in the field.

The same applies to mathematical properties. For example, "an indicator of average performance is consistent if adding the same publication to two different but equally large sets of publications never changes the way in which the indicator ranks the sets of publications relative to each other" (Waltman, van Eck, van Leeuwen, Visser, & van Raan, 2011, p. 43). Another mathematical property is homogeneous normalization. An indicator of average performance has the property of homogeneous normalization "if, in the case of a set of publications that all belong to the same field, the indicator equals the average number of citations per publication divided by the field's expected number of citations per publication" (Waltman et al., 2011, p. 43).

# 4   Methods of field-categorization for generating normalized indicators in bibliometrics

The field-categorization scheme which is used for performing field-normalization is an important point in assessing field-normalized indicators. If the field-categorization scheme does not represent fields validly, the calculated indicators can be questioned.

A basic property is also that the field normalization follows disciplinary stratification in science. Science is structured along disciplinary boundaries. Although interdisciplinary research is done in certain topics of nearly all disciplines, the disciplinary stratification exists nonetheless. "Academic science is divided into disciplines, each of which is a recognised domain of organised teaching and research. It is practically impossible to be an academic scientist without locating oneself initially in an established discipline" (Ziman, 1996, p. 69). The results of Mutz, Bornmann, and Daniel (2015) to the grant peer review process of the Austrian Science Fund (FWF) demonstrate that "cross-disciplinary research tends to be the exception rather than the rule in the grant proposals examined" (p. 35). It is the aim of field-normalization to consider different publication and citation cultures in science evaluation. Whereas in biomedicine rather recent literature is cited and in general more literature is cited, it is less and rather older literature in other disciplines, such as agricultural sciences (Marx & Bornmann, 2015). Also, within sub-disciplines, such as in economics, different publication and citation cultures exist and performing field-normalization is well advised (Bornmann & Wohlrabe, 2017). The impact of papers should be assessed within disciplinary boundaries to facilitate evaluation under fair conditions.

Methods of field-categorization can be classified based on two dimensions: (1) journal-based versus publication-based field definitions and (2) intellectual (i.e., expert-based) versus algorithmic field definitions (see overviews in Bornmann, Marx, & Barth, 2013; Waltman, 2016). Journal sets (as defined by Clarivate Analytics in Web of Science or Elsevier in Scopus) reflect disciplinary stratification, because most journals are developed within disciplines and an important characteristic of emerging new fields is the introduction of new journals. Intellectual field classifications of single publications by experts are restricted to single (broad) disciplines and reflect field stratification within single disciplines, because the field categorization schemes have been developed within the disciplines.

However, the use of citation relations and co-citations based on single publications (Ruiz-Castillo & Waltman, 2015) questions this basic requirement. The resulting clusters of publications cannot be clearly interpreted as fields (Haunschild, Schier, Marx, & Bornmann, 2018). "Because these 'fields' are algorithmic artifacts, they cannot easily be named (as against numbered), and therefore cannot be validated. Furthermore, a paper has to be cited or contain references in order to be classified, since the approach is based on direct citation relations. However, algorithmically generated classifications of journals have characteristics very different from content-based (that is, semantically meaningful) classifications" (Leydesdorff & Milojević, 2015, p. 201). The normalization on the basis of citation relations and co-citations might lead to highly correlated values with field-normalized scores based on journal sets or human field assignments. However, the clusters formed representing abstract fields are not plausibly communicable to the evaluated entities.

## 5 Werther effect for explaining the success of some indicators

Scientists produce contributions to the archive of knowledge in an exchange with other scientists and by formally referencing their literature. By publishing papers, scientists are perceived by other scientists. They receive a sense of their importance in the community from these perceptions. It is fair to say that scientists do not play any important role in their community if they are not present on the stage of awareness of other scientists (Franck, 2002). Since perception by other scientists is an important object for scientists, impact measurement is seen as a relevant instrument in research evaluation (and as a proxy of quality). Thus, many scientists undertake citizen bibliometrics to investigate their own performance and that of their colleagues. Unfortunately, in many analyses inadequate indicators are used for cross-field comparisons, such as the h index or journal impact factor (JIF, Garfield, 1999), or field-normalized indicators which are not part of the toolbox of professional bibliometricians, such as the RCR indicator described in the following. In the foundational years of bibliometrics, all bibliometrics were "citizen" bibliometrics (Derek de Solla Price and Eugene Garfield were neither of them "professional", yet set the course of the field for decades to come). However, the field of bibliometrics is so professionalized meanwhile (with its own journals and conferences) that professional and citizen bibliometricians can be separated more or less clearly.

The current use of indicators in research evaluation sometimes seems to follow what is known as the Werther effect in sociology. This effect has been introduced for explaining increasing rates of suicides in a region. It describes a situation in which a spike of emulation suicides follows a widely publicized suicide – following Goethe's novel "The Sorrows of Young Werther". "The more publicity devoted to a suicide story, the larger the rise in suicides" (Phillips, 1974, p. 340). The Werther effect in bibliometrics means that indicators are used by (citizen) bibliometricians, not because they have been developed and recommended by professional bibliometricians, but because they have been prominently published. The Werther effect can be demonstrated by the introduction of the RCR indicator. The RCR indicator was introduced in the reputable journal *PLOS Biology* as a newly designed field-normalized indicator (Hutchins, Yuan, Anderson, & Santangelo, 2016). Comments by Bloudoff-Indelicato (2015) and Naik (2016) followed in the similar reputable journal *Nature*. It seems that the indicator will be used in research evaluation (see https://icite.od.nih.gov), despite critical comments in the literature (Bornmann & Haunschild, 2017; Janssens, Goodman, Powell, & Gwinn, 2017; Waltman, 2015). For example, the RCR is included as field-normalized indicator in the new Dimensions platform developed by Digital Science (see https://www.dimensions.ai).

## 6 Conclusions

In order to negate the Werther effect in the application of bibliometric indicators, it is important that bibliometricians develop and analyze impact indicators against the backdrop of Popper's critical rationalism. Situations should be created or found in empirical research in which the measurement with new indicators can fail. An indicator is regarded as preliminarily valid if these situations could not be found or realized. Such an indicator should gain general acceptance in citizen and professional bibliometrics. Furthermore, studies investigating indicators should publish the results in such a way that they can be included in meta-analyses. The results of meta-analyses give guidance as to which indicators can be used as standard in bibliometrics.

Our proposal for an analysis of impact indicators against the backdrop of Popper's critical rationalism does not mean that we ask for a single (field-normalized) indicator which should be used in all evaluation contexts. Evaluations need various indicators that are tailored specifically to certain contexts. Standardization runs the risk of being based too much on a one-size-fits-all idea. It works against pluralism and innovation in bibliometrics. Having a variety of indicators also works against gaming of indicators. When there is only a single indicator that is used everywhere, researchers will have a strong incentive to try to game this indicator (Rijcke, Wouters, Rushforth, Franssen, & Hammarfelt, 2016).

We propose, however, that the various (field-normalized) indicators in the different contexts are continuously analyzed to their validity – against the backdrop of Popper's critical rationalism. Further, the results for each analyzed indicator from different studies should be summarized in meta-analyses. Then, a set of reliable and valid bibliometric indicators can be generated which includes recommended indicators from the professional field of bibliometrics for the use within and outside the field. Since revisions of existing indicators and new indicators should be done continuously (Hicks, Wouters, Waltman, de Rijcke, & Rafols, 2015), this set should always remain in the work-in-progress status. The generation and continuous revision of the set could be handled by the International Society for Informetrics and Scientometrics (ISSI) – the international association of scholars and professionals active in the interdisciplinary study science of science, science communication, and science policy (see http://www.issi-society.org).

With our appeal for standardization, we follow earlier initiatives in scientometrics. In the mid-1990s, Glänzel and Schöpflin (1994) saw the field of scientometrics in a deep crisis: "subfields are drifting apart, the field is lacking consensus in basic questions and of internal communication, the quality of scientometric research is questioned by other disciplines" (p. 375). To overcome the situation, they pled – among other things – for technical and scientific standards in research and publication of bibliometrics (see also Glänzel, 1996). About at the same time, Vinkler (1996) critically summarized the stage of indicators development in scientometrics as follows: "the indicators presented by different authors are mostly incompatible, impedes the development of the discipline and makes scientists, science managers and research policy makers mistrustful of the results of this discipline. Standardization of data, methods and indicators and that of their presentation are, therefore, urgently needed" (p. 238). The appeal for standardization by Vinkler (1996) was formulated based on his observation that many indicators have been constructed by researchers for a single study only.

Vinkler (1996) proposed as a solution for the problem of indicators' standardization that an international committee (organized by the ISSI) produces a "Manual of Recommended Scientometric Standards". An important requirement for the standardization of indicators is the systematic recording of the definitions of proposed or used indicators in published studies and reports. For example, Bornmann (2014) searched the literature for identifying definitions of highly-cited papers in bibliometrics. He found the following: "With definitions that relate to an absolute number, either a certain number of top cited papers (58%) or papers with a minimum number of citations are selected (17%). Approximately 23% worked with percentile rank classes" (p. 166). Comprehensive reviews of existing indicators can be found, for example, in the handbook by Todeschini and Baccini (2016) and the report by Rehn, Kronman, and Wadskog (2007).

In addition to a "Manual of Recommended Scientometric Standards" as proposed by Vinkler (1996), the programming languages of statistical software packages could be used to automatically produce bibliometric reports based on established standard indicators. The program Stata (StataCorp., 2017), for instance, introduced recently the command putdocx which creates Word documents with embedded Stata results using a given dataset. Based on putdocx, Bornmann and Haunschild (2018) developed the command bibrep.ado for the bibliometrics area. bibrep.ado can be used to produce a bibliometric report on single scientists (in Word format) based on a standardized set of bibliometric data. We could imagine that the ISSI publishes not only a manual including definitions of standard indicators as proposed by Vinkler (1996), but also commands, such as bibrep.ado, which could be used by professional or citizen bibliometricians to produce standardized bibliometric reports for different units (e.g. researchers, institutions, and countries).

# Acknowledgements

The authors thank Ludo Waltman for very helpful comments.